%% file: draft_D0tokpi0munu_resub_PRL_v3.tex
\documentclass[aps,prl,twocolumn,showpacs,amsmath,amssymb]{revtex4-2}
\usepackage{amsmath}
\usepackage{graphicx}
\usepackage{subfigure}
\usepackage{epstopdf}
\usepackage{color}
\usepackage{multirow}
\usepackage{setspace}
\usepackage{overpic}
\usepackage{amssymb}
\usepackage[bookmarksnumbered, pdfstartview=FitH,colorlinks,urlcolor=blue, citecolor=blue,linkcolor=blue] {hyperref}
\usepackage{lineno}
\usepackage{bm}
\usepackage{rotating}
\usepackage[utf8]{inputenc}

\let\oldequation\equation
\let\oldendequation\endequation

\renewenvironment{equation}
  {\linenomathNonumbers\oldequation}
  {\oldendequation\endlinenomath}

\begin{document}

\title{{\bf \boldmath Test of Lepton Universality and Measurement of the Form Factors of $D^0\to K^{*}(892)^-\mu^+\nu_\mu$ }}

\input{authorlist_2023-12-13}


\begin{abstract}
One of the main goals of studying semileptonic decays in flavor physics is to gain a better understanding of hadronic transitions in the nonperturbative region of Quantum Chromodynamics. This involves measuring Cabibbo-Kobayashi-Maskawa matrix elements, understanding form factors, and comparing them with theoretical predictions. We report a  first study of the semileptonic decay $D^0\rightarrow K^-\pi^0\mu^{+}\nu_{\mu}$  by analyzing an $e^+e^-$ annihilation data sample of  $7.93~\mathrm{fb}^{-1}$ collected at the center-of-mass energy of 3.773 GeV with the BESIII detector. The absolute branching fraction of $D^0\to K^-\pi^0\mu^{+}\nu_{\mu}$ is measured for the first time, providing necessary input for extracting the $c\rightarrow s$ Cabibbo-Kobayashi-Maskawa matrix element  through this process. This measurement allows us to test lepton flavor universality, with no indication of violation found.  Furthermore, a series of hadronic form factors have been determined and compared with theoretical predictions, which will impose stricter constraints on theoretical models.

\end{abstract}

\maketitle

\oddsidemargin  -0.2cm
\evensidemargin -0.2cm
Studies of the semileptonic (SL) decays of charmed mesons provide an ideal laboratory to explore the weak and strong interactions in mesons composed of heavy quarks~\cite{pr494_197}.  The SL partial decay width is related to the product of the hadronic form factors (FFs) describing the strong interactions in the initial and final hadrons, including non-perturbative effects of Quantum Chromodynamics (QCD), and a Cabibbo-Kobayashi-Maskawa (CKM) matrix element~\cite{prl10_531, ptp49_652}. Thus, precise measurements of FFs are important for testing the theoretical predictions where decay rates are calculated from first principles~\cite{prd52_2783,prd44_3567}.  
It is worth mentioning that in the study of hadronic transition FFs in the $D\rightarrow V \ell \nu_\ell $ decays, where $V$ refers to a vector meson, these transitions are notoriously difficult to model due to theoretical complexity. Therefore, for the decay $D^0\rightarrow K^{*}(892)^-\ell^+\nu_\ell$, different theoretical models provide varying predictions for the FFs~\cite{fpb14_66401,Ijmp21_6125-6172,cqm_2000,lfqm_2012,hmt_2005}, making experimental input indispensable for advancing our understanding. However, the FF measurements have only been performed in the $K^*(892)^-\to \bar{K}^0\pi^-$ decay mode, and the measurement precision is poor for the muonic decay channel ($\ell^+=\mu^+$)~\cite{plb_67-77}. In light of these challenges, it is interesting to study the dynamics of $D^0\rightarrow K^{*}(892)^-\mu^+\nu_\mu$ in a different decay mode, such as $K^*(892)^-\to K^- \pi^0$, which will provide more precise FF results and  a more rigorous validation for theoretical calculations.

In the standard model (SM), SL $D$ decays   also offer an excellent opportunity to test  lepton flavor universality~(LFU)~\cite{ARNPS73_285-314,NSR8_181,prd91_094009,cpc_063107}. References~\cite{prd91_094009,cpc_063107} note that there may indeed be observable LFU violation effects in SL decays mediated via $c\rightarrow s \ell \nu$.
In particular, for the $D\rightarrow V \ell \nu_\ell $ decays, the multiple polarization states of vector mesons provide more physical information, allowing for a more detailed examination of the differences between different types of leptons. This helps to better understand the mechanisms of weak interactions and potential new physics effects, making the testing of LFU violations more reliable and significant. Theoretical studies have shown great interest in the $D^0\rightarrow K^{*}(892)^-\ell^+\nu_\ell$ decay process and the ratio $\frac{{\mathcal B}(D^0\to K^{*}(892)^-\mu^+\nu_\mu)}{{\mathcal B}(D^0\to K^{*}(892)^-e^+\nu_e)}$ is predicted to be $0.92-0.99$~\cite{fpb14_66401,Ijmp21_6125-6172,cpc_063107,prr2_0431129,prd92_054038,prd96_016017}.  There are significant differences among various theoretical models~\cite{prr2_0431129, fpb14_66401,prd96_016017} in the predicted branching fraction~(BF) results for $D^0\rightarrow K^{*}(892)^-\mu^+\nu_\mu$, varying in the range ($1.93-3.09$)\%. Hence, it is important to measure the $D^0\to K^{*}(892)^-\mu^+\nu_\mu$ decay with high precision to distinguish among these theories and also test $\mu-e$ LFU.

In this Letter, we report the first measurement of the absolute BF for the $D^0\rightarrow K^{-}\pi^0\mu^+\nu_\mu$ decay and investigate its dynamics. When measuring the absolute BF of the $D^0\rightarrow K^-\pi^0\mu^{+}\nu_{\mu}$ decay, the dominant component arises from the decay chain $D^0\rightarrow K^{*}(892)^-\mu^+\nu_\mu$, $K^{*}(892)^-\rightarrow K^-\pi^0$. The process involving the intermediate $K^{*}(892)^-$ meson is referred to as the $P$ wave, while the direct transition without the  $K^{*}(892)^-$ meson is denoted as the $S$ wave in this analysis. These measurements are performed using an $e^+e^-$ annihilation data sample corresponding to an integrated luminosity of $7.93~\mathrm{fb}^{-1}$~\cite{Lum_1,Lum_2,Lum_3} produced at  the center-of-mass energy of $\sqrt{s}=3.773$ GeV with the BEPCII collider and collected with the BESIII detector.  Charge-conjugate modes are implied throughout this Letter.

A description of the design and performance of the BESIII detector can be found in Refs.~\cite{Ablikim:2009aa,detector}.  The inclusive Monte Carlo (MC) sample, described in Refs.~\cite{geant4,kkmc,evtgen,pdg16,lundcharm,photos},  is used to determine the selection efficiencies and estimate backgrounds. In the generation of simulated signals  $D^0\rightarrow K^{-}\pi^0\mu^+\nu_\mu$, we take into account the knowledge of the FFs obtained in this work. 

The analysis utilizes both “single-tag” (ST) and “double-tag” (DT) samples of $D$ decays.  At $\sqrt{s}=3.773$ GeV, the $\psi(3770)$ resonance decays mainly into a $D\bar{D}$ pair. If a $\bar{D}$  meson is fully reconstructed, the presence of a  $D$ meson is guaranteed. Thus, in the system recoiling against a $\bar{D}$ meson, the $D^0\rightarrow K^-\pi^0\mu^{+}\nu_{\mu}$ candidate can be selected. The ST samples are reconstructed  in six hadronic final states listed in Table~\ref{tab:numST}, which are referred to as the tag modes. A subset of events is  selected within each ST sample, where the other tracks in the event correspond to the decay $D^0\rightarrow K^{-}\pi^0\mu^+\nu_\mu$, denoted as the DT sample.
The BF for the $D^0\rightarrow K^{-}\pi^0\mu^+\nu_\mu$ decay is expressed as
\begin{equation}
\mathcal{B}_{\rm SL}=\frac{N_{\rm DT}}{N^{\rm tot}_{\rm ST}\cdot\bar\epsilon_{\rm SL}}, \label{eq:branch}
\end{equation}
where $N_{\rm DT}$ and $N^{\rm tot}_{\rm ST}=\sum^{6}_{i=1} N^{i}_{\rm ST}$   are the total DT and ST yields  summing over tag mode $i$, $\bar\epsilon_{\rm SL}=\sum^{6}_{i=1} \frac{N^{i}_{\rm ST}}{N^{\rm tot}_{\rm ST}}\cdot\frac{\epsilon^i_{\rm DT}}{\epsilon^i_{\rm ST}}$
is the averaged efficiency of reconstructing the $D^0\rightarrow K^{-}\pi^0\mu^+\nu_\mu$ decay,  where $\epsilon^i_{\rm ST}$ and  $\epsilon^i_{\rm DT}$ 
are the ST and DT efficiencies for the $i$th tag mode, respectively.

\begin{figure}[tp!]
\begin{center}
   \flushleft
   \begin{minipage}[t]{8.0cm}
\includegraphics[width=\linewidth]{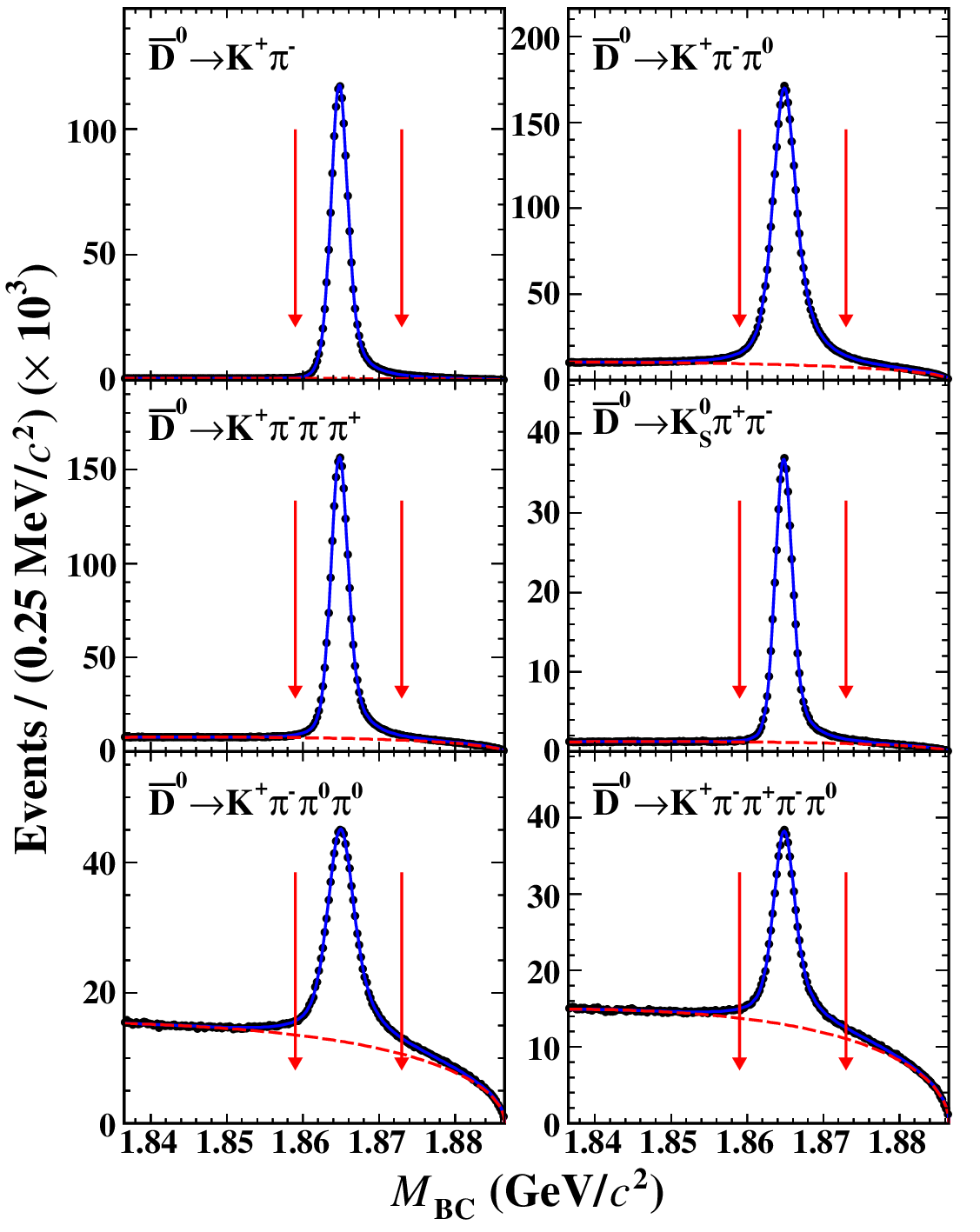}
   \end{minipage}
\caption{ Fits to the $M_{\rm BC}$ distributions of the ST $\bar{D}^0$ candidates. The points are data,
the blue curves are the best fits, and the red dashed curves are the fitted combinatorial background shapes.  The pair of red
arrows show the $M_{\rm BC}$ signal window of (1.859, 1.873) GeV/$c^2$.}
\label{fig:tag_md0}
\end{center}
\end{figure}

A detailed description of the selection criteria for  $\pi^\pm$, $K^\pm$,  $\gamma$, $K^0_S$, and $\pi^0$ candidates for the ST is provided in Refs.~\cite{prl121_171803, prl123_231801}. 
Two variables are calculated for each possible ST candidate:   the beam-constrained mass $M_{\rm BC}\equiv\sqrt{E_{\mathrm{beam}}^{2}/c^4-|\vec{p}_{\bar D^0}|^{2}/c^2}$ and  the energy difference $\Delta E\equiv E_{\bar D^0}-E_{\mathrm{beam}}$, where $E_{\mathrm{beam}}$ is the beam energy, and $E_{\bar D^0}$ and $\vec{p}_{\bar D^0}$ are the total energy and momentum of the ST $\bar{D}^0$ meson candidate in the $e^+e^-$ center-of-mass frame, respectively. For each ST mode, if there are multiple candidates in an event, only the one with the smallest $|\Delta E|$ is kept for further analysis.  The $\Delta E$ requirements and ST efficiencies are summarized in Table~\ref{tab:numST}.
The tagged decay yields are determined separately for the six tag channels. The yields are obtained by fitting the signal and background contributions to the $M_{\rm BC}$ distribution (Fig.~\ref{fig:tag_md0}) of the events passing the $\Delta E$ cuts. The signal shape is modeled by the MC simulated shape convolved  with a double Gaussian function, while the background shape is described by the ARGUS function~\cite{plb241_278}. The yields (listed in Table~\ref{tab:numST}) are determined by subtracting the numbers of background events from the total numbers of events in the $M_{\rm BC}$ signal region. Summing over all ST modes, we obtain the total ST yield  $N^{\rm tot}_{\rm ST}=(7895.8\pm3.4)\times10^{3}$.

\begin{table}[tp!]
\caption{ The $\Delta E$ requirements, the ST efficiencies, $ \epsilon^{i}_{\rm ST}$, and the obtained ST $\bar{D}^0$ yields in the data, $N^{i}_{\rm ST}$. The efficiencies do not include the BFs for  $K_{S}^0 \to \pi^+\pi^- $ and  $\pi^0 \to \gamma\gamma$. The uncertainties are statistical only.}
\begin{center}
\scalebox{0.90}{
\begin{tabular}
{lccr} \hline\hline Decay mode      & $\Delta E$ (GeV)        & $ \epsilon^{i}_{\rm ST}~(\%)$     &  $N^{i}_{\rm ST}$ ($\times 10^3$)      \\
       	\hline
		$ K^+ \pi^-$                   & ($-0.027, 0.027$) & 65.34 $\pm $ 0.01 & 1449.3  $\pm$ 1.3 \\
		$K^+ \pi^- \pi^0$              & ($-0.062, 0.049$) & 35.59 $\pm $ 0.01 & 2913.2  $\pm$ 2.0 \\
		$ K^+ \pi^+ \pi^- \pi^-$       & ($-0.026, 0.024$) & 40.83 $\pm $ 0.01 & 1944.2  $\pm$ 1.6\\
		$ K_S^0 \pi^+ \pi^-$           & ($-0.024, 0.024$) & 37.49 $\pm $ 0.01 &  447.7  $\pm$ 0.7\\
		$K^+ \pi^- \pi^0 \pi^0$        & ($-0.068, 0.053$) & 14.83  $\pm $ 0.01 &  690.6  $\pm$ 1.3\\
		$ K^+ \pi^+ \pi^- \pi^- \pi^0$ & ($-0.057, 0.051$) & 16.17 $\pm $ 0.01 &  450.9  $\pm$ 1.1\\
\hline\hline
\end{tabular}
}
\label{tab:numST}
\end{center}
\end{table}

Candidates for the $D^0\rightarrow K^{-}\pi^0\mu^+\nu_\mu$ decay  are selected from the remaining tracks recoiling against the ST $\bar{D}^0$ mesons. Candidates for $K^{-}$ and $\pi^0$ are identified with the same  criteria as  those used in the ST selection. A kinematic fit, constraining  $M_{\gamma\gamma}$ to the known $\pi^0$  mass~\cite{pdg16}, is performed.  In the case of multiple $\pi^0$ candidates, only the one with the smallest  $\chi^{2}$ is retained.  
Particle identification (PID) for the $\mu^+$ includes the measurements of the specific ionization energy loss by the  drift chamber, the flight time by the time-of-flight system, and the energy deposited in the electromagnetic calorimeter (EMC). Based on these measurements, we calculate the combined confidence levels for positron ($\mathcal{L}_e$), muon ($\mathcal{L}_\mu$), pion ($\mathcal{L}_\pi$), and kaon ($\mathcal{L}_K$) hypotheses for each charged track. The muon candidate must satisfy $\mathcal{L}_\mu > \mathcal{L}_\pi$,  $\mathcal{L}_\mu > \mathcal{L}_K$,  $\mathcal{L}_\mu > \mathcal{L}_e$, $\mathcal{L}_{\mu} > 0.001$ and $E_{\rm EMC} \in (0.10, 0.28) $~GeV, where $E_{\rm EMC}$ represents the energy deposition in the EMC. The two requirements $\mathcal{L}_\mu > \mathcal{L}_\pi$ and $E_{\rm EMC}$ together suppress about 85\% of background at the cost of 46\% of signal.
  To  suppress backgrounds from hadronic $D$ decays, the maximum energy of any photon that is not used in the SL selection, $E^{\rm max}_{\rm extra~ \gamma}$, is required to be less than 0.25~GeV, and the number of extra unused charged tracks, $N^{\rm char}_{\rm extra}$, and extra $\pi^0$ from two unused photons, $N^{\pi^{0}}_{\rm extra}$, must both be zero.   Here, the selection criteria for photon, charged tracks, and $\pi^0$ are consistent with Refs.~\cite{prl121_171803, prl123_231801}.
 
The energy $E_{\rm miss}$ and momentum $\vec{p}_{\rm miss}$ of the missing neutrino are reconstructed using energy and momentum conservation. They are calculated by $E_{\rm miss}\equiv E_{\mathrm{beam}}-\sum_{j}E_{j}$ and $\vec{p}_{\rm miss} \equiv \vec{p}_{D^0}-\sum_{j}\vec{p}_{j}$ in the initial $e^+e^-$ rest frame.
The index $j$ sums over the $K^-$, $\pi^0$ and $\mu^+$ of the signal candidate, and $E_j$ and $\vec{p}_j$ are the energy and momentum of the $j$th particle, respectively.
The $D^0$ momentum is given by
$\vec{p}_{D^0} \equiv -\hat{p}_{\bar{D}^0}\sqrt{E_{\mathrm{beam}}^2/c^2-m^2_{\bar{D}^0}c^2},$
where $\hat{p}_{\bar{D}^0}$ is the
momentum direction of the ST $\bar{D}^0$ and $m_{\bar{D}^0}$ is the known  $\bar{D}^0$ mass~\cite{pdg16}. The presence of the undetected neutrino is inferred by using the variable $U_{\rm miss}$ defined by
\begin{equation}
U_{\rm miss} \equiv E_{\rm miss}-|\vec{p}_{\rm miss}|c .
\end{equation}
The potential background from $D^0\rightarrow K^-\pi^+\pi^0$ is suppressed by the requirements of $M_{K^{-}\pi^0\mu^+} <$ 1.80~GeV/$c^2$ and $U^{\prime}_{\rm miss}>0.04$ GeV, where $M_{K^{-}\pi^0\mu^+}$ is the $K^{-}\pi^0\mu^+$ invariant mass and  $U^{\prime}_{\rm miss}$ is  defined $U^{\prime}_{\rm miss} \equiv E^{\prime}_{\rm miss}-|\vec{p~}^{\prime}_{\rm miss}|c$. 
Here, $E^{\prime}_{\rm miss}\equiv E_{\mathrm{beam}}-\sum_{j}E_{j}$ and  $\vec{p~}^{\prime}_{\rm miss} \equiv \vec{p}_{D^0}-\sum_{j}\vec{p}_{j}$,  and $j$ only sums over the $K^-$ and $\mu^+$ candidates of the signal candidate. To suppress the background  from $D^0\to K^-\pi^+\pi^0\pi^0$,  the opening angle between the missing momentum and the most energetic unused shower when found, $\theta_{\vec{p}_{\rm miss}, \gamma}$, is required to satisfy $\cos\theta_{\vec{p}_{\rm miss}, \gamma} <$ 0.48.

\begin{figure}[tp!]
\begin{center}
   \flushleft
        \includegraphics[width=\linewidth]{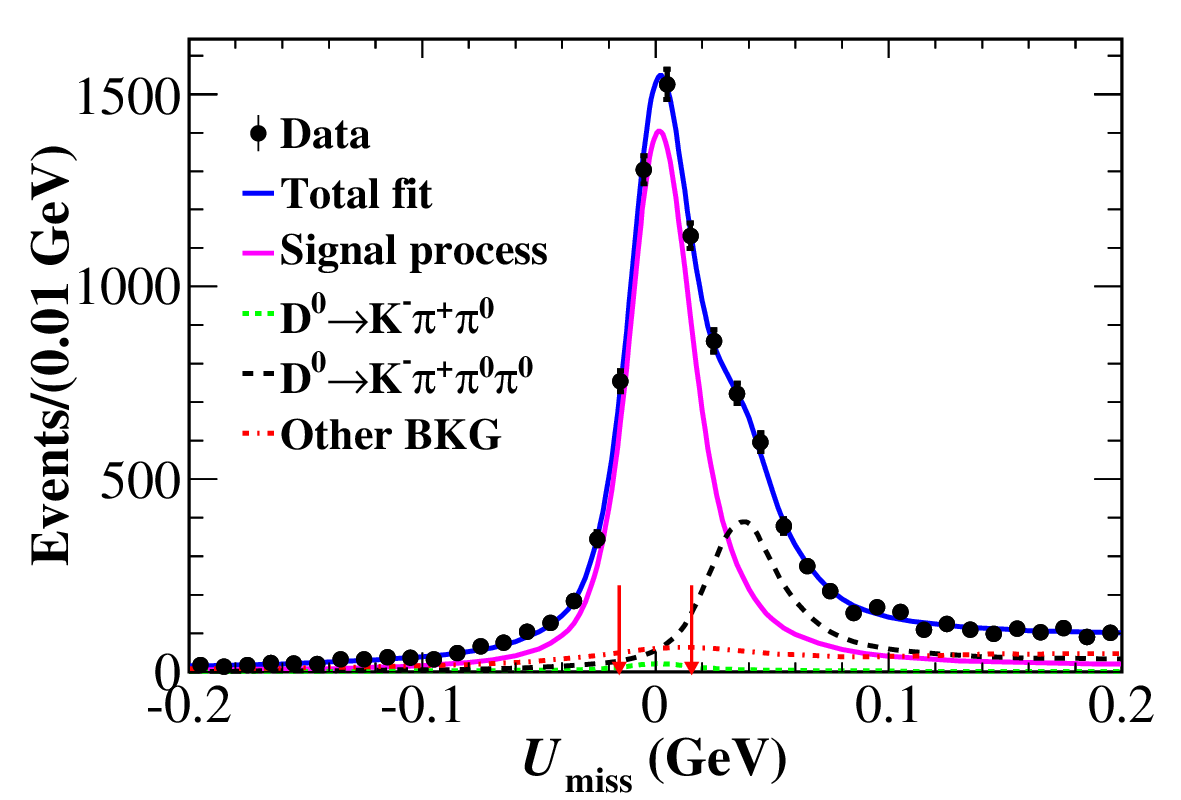}
   \caption{ Fits to $U_{\rm miss}$  distributions of the candidate events for $D^{0}\to K^-\pi^{0}\mu^{+}\nu_{\mu}$. 
Points with error bars represent data. The blue solid curve denotes the total fit, and the violet solid  curve shows the signal process.
Green short-dashed, black  long-dashed, and  dash-dotted red curves  are the background contributions from $D^0\to K^-\pi^+\pi^0$, $D^0\to K^-\pi^+\pi^0\pi^0$,  and  the other sources, respectively.  The pair of red arrows show the signal window of $|{ U_{\rm miss}}|<0.015$ GeV used for the amplitude analysis.}
\label{fig:bf}
\end{center}
\end{figure}

After imposing all above selection criteria, the resulting $U_{\rm miss}$ distribution of the accepted candidates is exhibited in Fig.~\ref{fig:bf}.
To obtain the signal yield, an unbinned maximum likelihood fit to the $U_{\rm miss}$ distribution is performed.
In the fit, the signal is modeled by the MC-simulated shape convolved with a Gaussian function with free parameters. The peaking background from $D^0\to K^-\pi^+\pi^0$  is  fixed according to the MC simulation and the yield of  the dominant background from $D^0\to K^-\pi^+\pi^0\pi^0$  is floated. Both of these background shapes are smeared with the same Gaussian function as used for the signal. 
Other backgrounds, mainly from open charm production and continuum $q\bar q$, are evaluated using the MC simulation.  The number of DT events is found to be $N_{\rm DT}=6436 \pm 119_{\rm stat}$.  The averaged detection efficiency $\bar\epsilon_{\rm SL}$ is estimated to be  $(11.25\pm0.02)\%$. The efficiency does not include the BFs for $K^*(892)^-\to K^- \pi^0$ and $\pi^0\to \gamma \gamma$~\cite{pdg16}. Using  Eq.~(\ref{eq:branch}), the corresponding BF is determined as $\mathcal B({D^0\rightarrow K^{-}\pi^0\mu^+\nu_\mu})=(0.733 \pm 0.014_{\rm stat})\%$. 

The systematic uncertainties in the BF measurement are discussed below. The efficiencies of $\mu^+$ and $K^{-}$ tracking, PID, and $\pi^0$ reconstruction are verified using $e^+e^-\to \gamma\mu^+\mu^-$ events and DT $D\bar D$ hadronic events, respectively. We assign the uncertainties of  $\mu^+$  tracking (PID), $K^{-}$ tracking (PID), and $\pi^0$ reconstruction to be 0.1\% (0.4\%), 0.2\% (0.1\%), and 0.7\%, respectively. The total uncertainty associated with the $E^{\rm max}_{\rm extra\gamma}$, $N^{\pi^{0}}_{\rm extra}$, and $N^{\rm char}_{\rm extra}$  requirements are estimated to be 0.5\%  by using the control sample of $D^0\to K^-\pi^{+}\pi^{0}$. The efficiency of the $U^{\prime}_{\rm miss}$ requirement is very high and the uncertainty is thus negligible.   
The uncertainty associated with the $M_{\rm K^-\pi^0\mu^{+}}$   requirement is estimated to be 0.6\% by analyzing the DT events of $D^0\to K^-\pi^{0} e^+\nu_e$. The uncertainty associated with the $\cos\theta_{\vec{p}_{\rm miss}, \gamma}$   requirement is estimated to be 0.5\% by analyzing the DT sample $D^0\to K_S^{0}\pi^-e^+\nu_e$. The uncertainty associated with the fit to the $U_{\rm miss}$ distribution is estimated to be 0.7\% by varying the signal shape, the combinational background shape and the quoted BF of the fixed $K^-\pi^+\pi^0$ peaking background by $\pm 1\sigma$. The uncertainty from the ST yield is determined to be 0.1\%~\cite{prd111_222}.   The uncertainty related to the signal MC model is estimated to be 0.5\% by comparing the DT efficiencies with variations in the input FF parameters by $\pm 1\sigma$. The uncertainty due to the limited MC statistics is 0.2\%. The uncertainties due to the quoted BF of  $\pi^0 \to \gamma \gamma$ is  0.1\%. The systematic uncertainty contributions are summed in quadrature  to obtain the total systematic uncertainty of 1.5\%.

To study the $K^-\pi^0$ system and measure the FF, we require $|{ U_{\rm miss}}|<0.015$ GeV to select samples with high purity for the amplitude analysis; this
leads to 3375 events with a background fraction  $f_b$ = (12.6~$\pm$~0.7)\%. The differential decay rate  can be expressed in terms of five kinematic variables~\cite{ pr168_1926,prd46_5040}: the square of the invariant mass of the $K^-\pi^0$ system ($M_{K^-\pi^0}^2$), the square of the invariant mass of the $\mu^+\nu_\mu$ system ($q^2$), the angle between the momentum of the $K^-$  in the $K^-\pi^0$  rest frame and the momentum of the $K^-\pi^0$ system in the $D^0$ rest frame ($\theta_{K^-}$), the angle between the momentum of the $\mu^+$  in the $\mu^+\nu_\mu$  rest frame and the momentum of the $\mu^+\nu_\mu$ system in the $D^0$ rest frame ($\theta_\mu$), and $\chi$, the angle between the normals of the decay planes defined in the $D^0$ rest frame by the $K^-\pi^0$ pair and the $\mu^+\nu_\mu$ pair.
The differential decay rate as a function of these variables is given in Ref.~\cite{prd46_5040}.  Modifications due to the nonzero muon mass are detailed in Ref.~\cite{cpc_063101}. For the amplitude of the $P$ wave, we use a Breit-Wigner function to describe $K^{*}(892)^-$ resonance. A barrier factor, $r_{\rm BW}$, related to the meson effective  radii is included in the decay amplitude, which is fixed to be 3.07 GeV$^{-1}$~\cite{prd94_032001}.
The helicity FFs can in turn be related to the two axial-vector FFs, $A_1(q^2)$ and $A_2(q^2)$, as well as the vector FF, $V(q^2)$. The $A_{1,2}(q^2)$ and $V(q^2)$ are all described as the simple pole form $A_{1,2}(q^2)=A_{1,2}(0)/(1-q^2/M^2_A)$ and $V(q^2)=V(0)/(1-q^2/M^2_V)$, with pole masses $M_V=M_{D_s^*(1^-)}=2.1$~GeV/$c^2$ and $M_A=M_{D_s^*(1^+)}=2.5$~GeV/$c^2$~\cite{pdg16}.  At $q^2=0$, the FF  ratios, $r_V=V(0)/A_1(0)$ and $r_2=A_2(0)/A_1(0)$,  are determined from the fit to the differential decay rate.
The $S$ wave parametrization is described in Refs.~\cite{prd94_032001,prd83_072001}, in which the scattering length, $a^{1/2}_{\rm S,BG}$, and the relative intensity, $r_S$, are determined by the amplitude analysis. The dimensionless coefficient $r_S^{(1)}$ and the effective range $b^{1/2}_{\rm S,BG}$ are fixed to the values obtained from the $D^+\to K^-\pi^+e^+\nu_e$ analysis~\cite{prd94_032001} analogously to what was done in Ref.~\cite{prd99_011103}.

The  amplitude analysis is performed using an unbinned maximum likelihood fit.
The negative log likelihood is defined as~\cite{prd85_122002}
\begin{equation} 
\begin{aligned}
 -\ln\!{\cal L} &=  -\sum_{i=1}^{N}\ln  \left[(1-f_{b}) \frac{\omega(\xi_{i},\eta)}{\int \omega(\xi_{i},\eta) \, \epsilon(\xi_{i}) \, R_4(\xi_{i}) \, d\xi_{i}} \right. \\
 &\quad \left. + f_{b}\frac{B_{\epsilon}(\xi_{i})}{\int B_{\epsilon}(\xi_{i}) \, \epsilon(\xi_{i}) \, R_4(\xi_{i}) \,  d\xi_{i} }\right],
  \label{eq:lnL}
\end{aligned}
\end{equation}
where $\xi_{i}$ denotes the five kinematic variables characterizing an event  and $\eta$ denotes the fit parameters  such as $r_V$ and $r_2$;
$\omega(\xi_{i},\eta)$ is the decay intensity, $B_{\epsilon}(\xi_{i})$ is defined to be the  background  distribution corrected by the acceptance function $\epsilon(\xi_{i})$, and $R_4(\xi_{i})$ is an element of four-body phase space. The background shape is modeled with the inclusive MC sample and its fraction $f_b$ is fixed according to the result of the $U_{\rm miss}$ fit.  The normalization integral in the denominator is calculated by MC integration~\cite{prd94_032001}.

\begin{figure}[tp!]
\begin{center}
   \flushleft
   \begin{minipage}[t]{8.8cm}
      \includegraphics[width=0.49\linewidth]{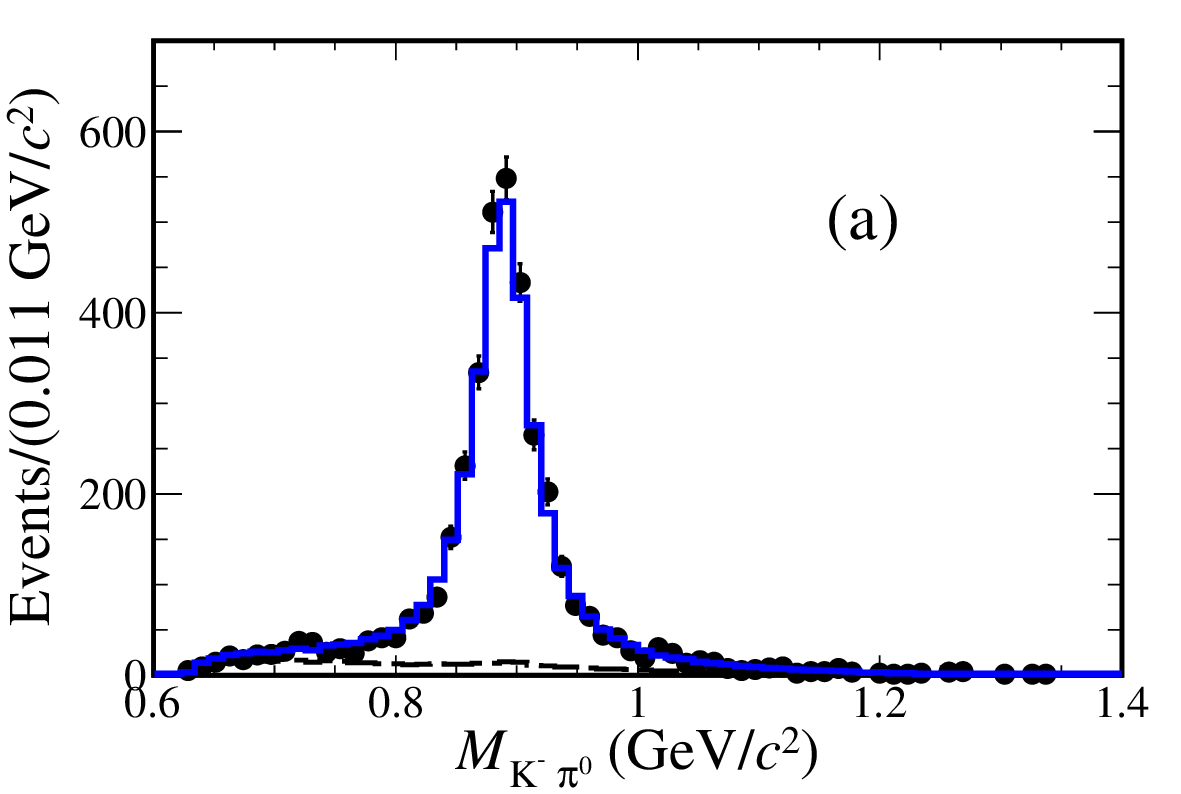}
      \includegraphics[width=0.49\linewidth]{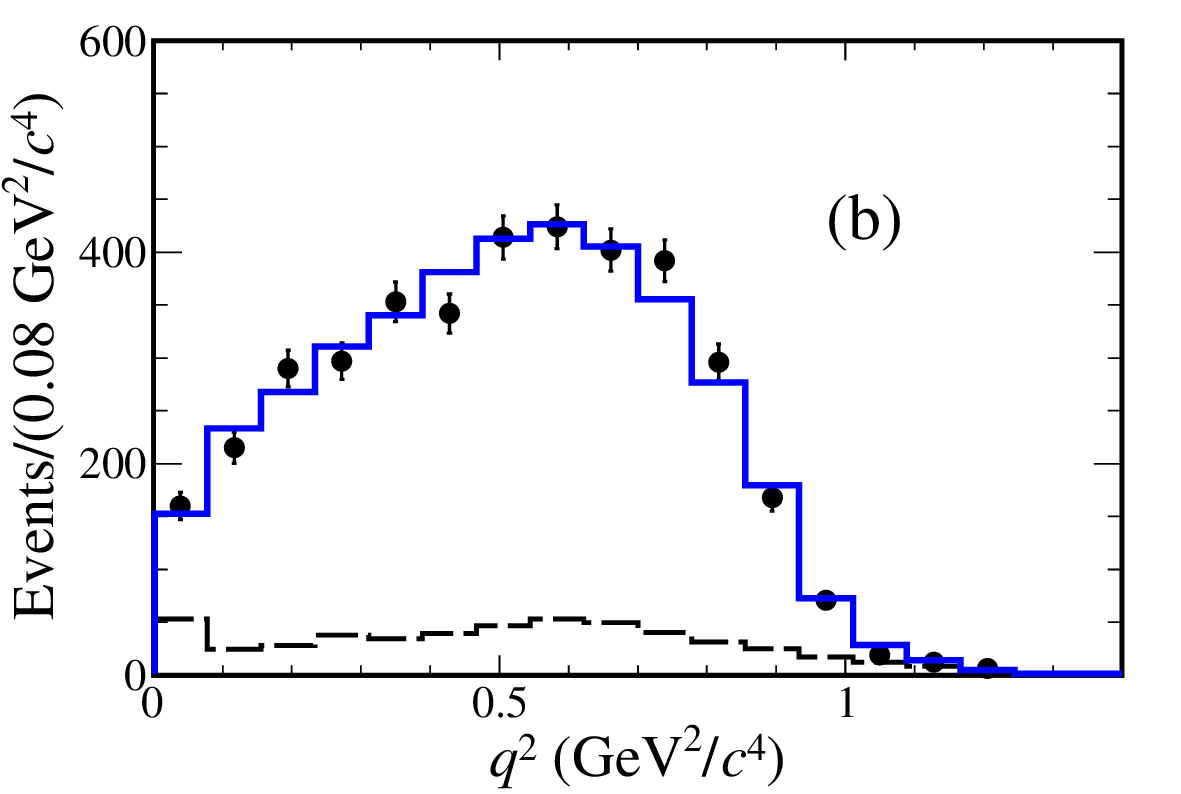}
       \includegraphics[width=0.49\linewidth]{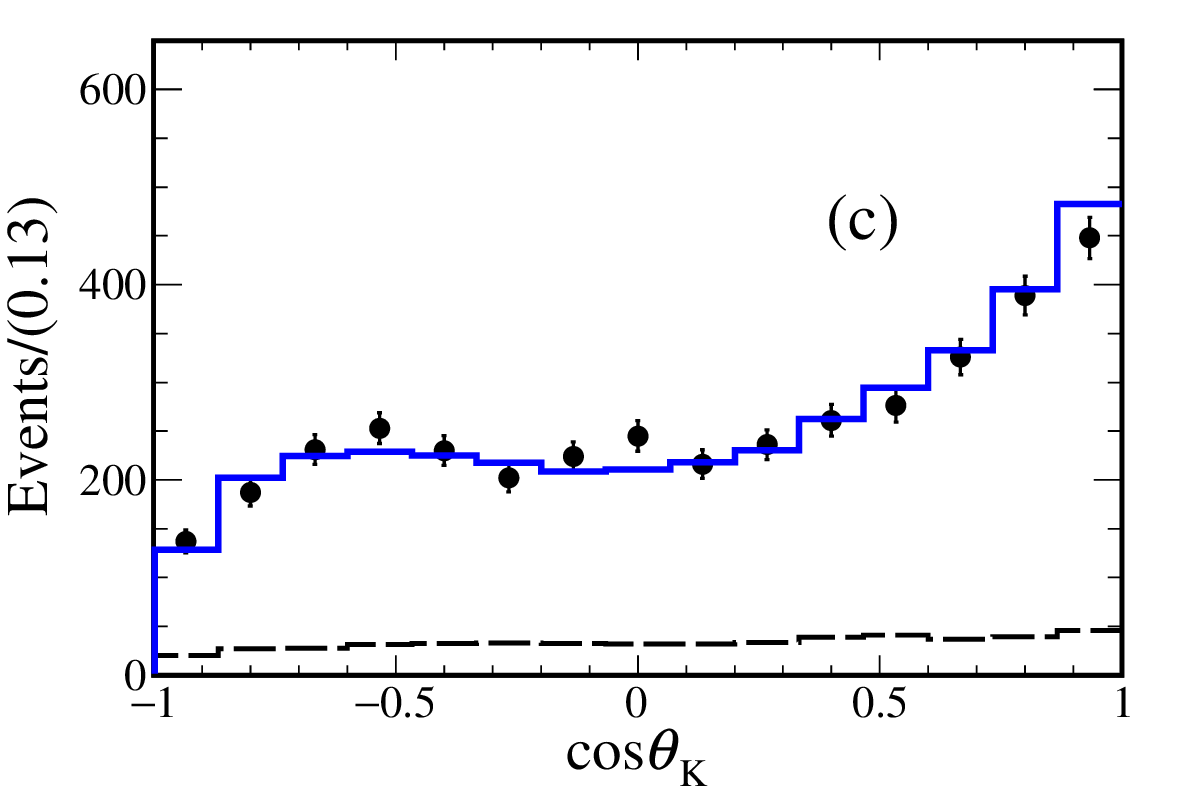}
      \includegraphics[width=0.49\linewidth]{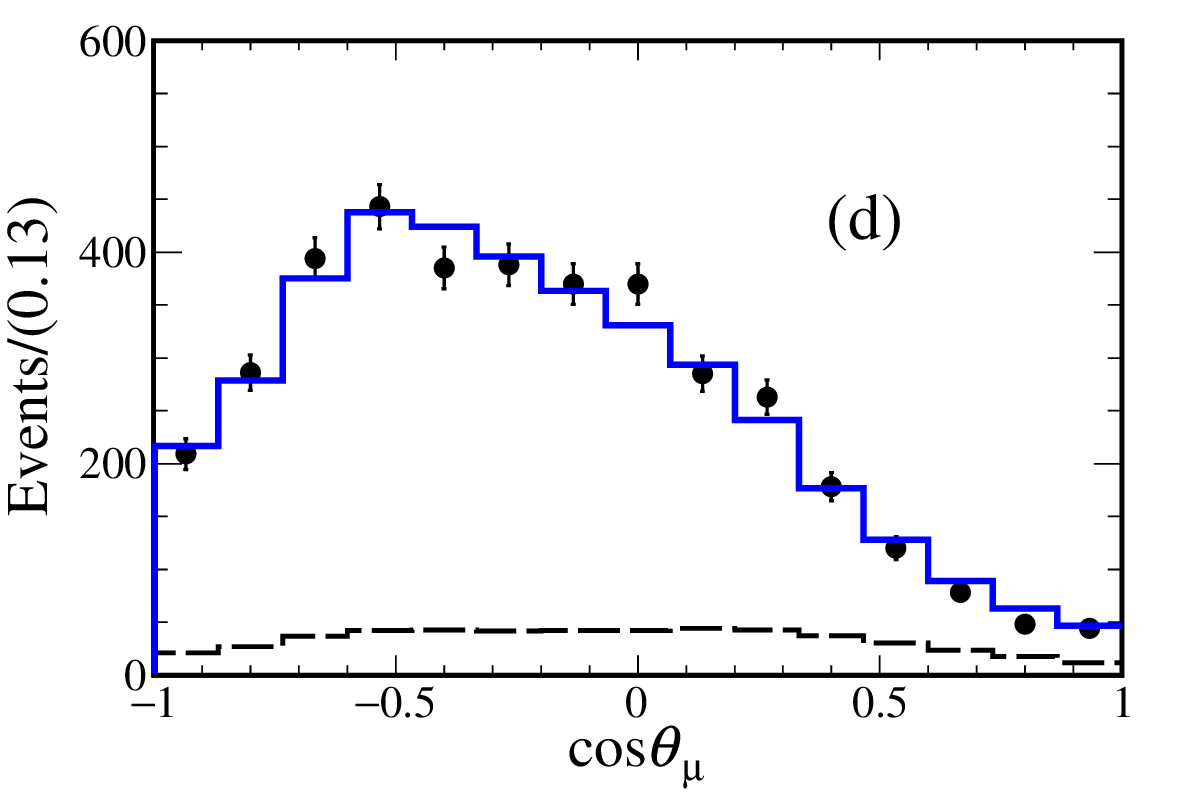}
      \includegraphics[width=0.49\linewidth]{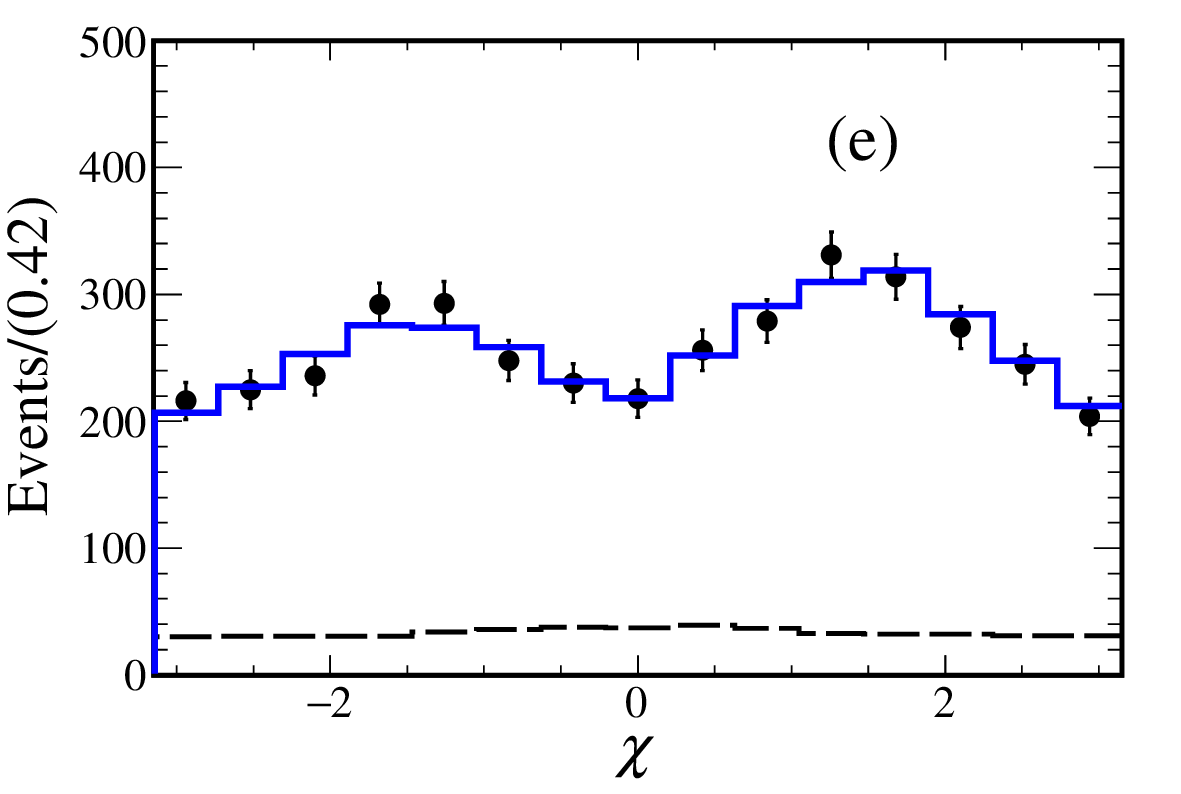}
   \end{minipage}
   \caption{Projections onto five kinematic variables  (a) $M_{K^-\pi^0}$, (b) $q^2$,  (c) $\cos\theta_K$,  (d) $\cos\theta_\mu$, and (e) $\chi$ for $D^0\rightarrow K^{-}\pi^0\mu^+\nu_\mu$. The dots with error bars are data,  the blue lines are the fit results, and the dashed lines show the sum of the simulated background contributions.}
\label{fig:formfactor}
\end{center}
\end{figure}

The nominal solution of the amplitude analysis fit, together with the fractions of both components, is summarized in Table~\ref{tab:FitResults}. The amplitudes of $S$ wave and $P$ wave are orthogonal, therefore the interference between them is 0. The projected distributions of the fit onto the fitted variables are shown in Fig.~\ref{fig:formfactor}. The possible contributions of the $K^*(1410)$ and $K^*_2(1430)$ are consistent with zero, and are hence ignored with current statistics. 

\begin{table}
\begin{center}
\caption{The fit results, where the first uncertainties are statistical and the second are systematic. } \normalsize
\begin{tabular}
{lc} \hline\hline  Variable~~~~~~~~~~~~~~~~~~~~~~~     &  ~~~~~~~Value                       \\ \hline
$r_V$                                                               &~~~~~~$1.37\pm0.09\pm0.03$     \\
$r_2$                                                               &~~~~~~$0.76\pm0.06\pm0.02$     \\
$M_{K^{*}(892)^-}$ (MeV/$c^2$)                                      &~~~$892.89\pm0.56\pm0.07$          \\
$\Gamma_{K^{*}(892)^-}$ (MeV)                                       &~~~~~$47.65\pm1.26\pm0.36$          \\
$r_S$ (GeV)$^{-1}$                                                  &~~~~$-8.40\pm0.34 \pm0.38$       \\
$a^{1/2}_{\rm S,BG}$ (GeV/$c$)$^{-1}$                               &~~~~~~$1.71\pm0.22\pm0.12$     \\
$f_{K^{*}(892)^-}(\%)$                                                     &~~~~~$94.24\pm0.35\pm0.29$     \\
$f_{S{\rm wave}}(\%)$                               &~~~~~~$5.76\pm0.35\pm0.29$     \\
\hline\hline
\end{tabular}
\label{tab:FitResults}
\end{center}
\end{table}

\begin{table*}
\begin{center}
\caption{Systematic uncertainties (in \%) of the fitted parameters. }
\begin{tabular}{lcccccccc}
\hline
\hline \normalsize
Source                                 & {\;\;$r_V$\;\;}  & {\;\;$r_2$\;\;} &{\;\;$r_S$\;\;} &$m_{K^{*}(892)^-}$&$\Gamma^0_{K^{*}(892)^-}$&$a^{1/2}_{\rm S,BG }$&$f_{K^{*}(892)^-}$&$f_{S {\rm wave}}$\\
\hline
{Background   fraction \;\;}              & 0.03   & 0.51 & 0.73 &  $<$ 0.01& 0.51& 1.13&  0.04 &0.72  \\
Background   shape                  &  0.06  &  0.13& 0.04 &  $<$ 0.01& 0.06&  0.06&  0.04 &0.12   \\

$r^{(1)}_{S}$                                             &  0.17  &  0.11& 4.14 &  ~~~0.01& 0.32&  1.06 & 0.26&  4.28\\
$b^{1/2}_{\rm S,BG}$                                     &  0.18  &  0.73& 0.42 & $<$ 0.01& 0.04&  6.90&  0.04&  0.59 \\
$r_{\rm BW}$                                                  &  0.31  &  0.38& 1.26 &  $<$ 0.01& 0.46&  0.22&  0.13&  2.23\\
$m_V$                                                      &  1.13  &  0.05& 0.05 & ~~~0.01& 0.01&  0.09&  0.01&  0.13 \\
$m_A$                                                      &  1.48  &  2.45& 0.66 &  $<$ 0.01& 0.01&  0.43&  0.02&  0.21 \\
Experimental effects                                 &  0.61 &  0.57&  0.49 &   $<$ 0.01& 0.11&  0.08& 0.04&  0.56 \\
\hline
Total                                                           & 2.00 & 2.71  & 4.48& ~~~0.01&  0.76&  7.09&  0.31 & 4.96\\
\hline
\hline
\end{tabular}
\label{tab:Syserr}
\end{center}
\end{table*}

The systematic uncertainties of the fitted parameters and the fractions of $S$ wave and $K^{*}(892)^-$ components are evaluated as the difference between the fit results in nominal conditions and those obtained after changing a variable or a condition by an amount corresponding to the estimated uncertainty of that quantity.
The systematic uncertainties due to the background  fraction and background shape requirements are estimated by varying the background fraction  $f_{b}$ by $\pm 1\sigma$ and  varying the cross section of the dominant background from $e^+e^- \to q\bar{q}$~\cite{corss_section}, respectively.  The systematic uncertainties in the fixed parameters of $r_S^{(1)}$, $b^{1/2}_{\rm S,BG}$,  and $r_{\rm BW}$ are estimated by varying their input values by $\pm 1\sigma$~\cite{prd99_011103}.  The systematic uncertainties in the fixed parameters of $m_V$ and $m_A$ are estimated by varying their input values by $\pm 100$~MeV/$c^2$~\cite{prd94_032001}. To estimate the systematic uncertainty of experimental effects related to the reconstruction efficiency, the fit is performed by varying the PID and  tracking efficiencies according to their uncertainties. These differences are assigned as the systematic uncertainties and summarized in Table~\ref{tab:Syserr}, where the total systematic uncertainties are obtained by adding all contributions in quadrature.

In summary, the SL decay $D^0\rightarrow K^-\pi^0\mu^+\nu_{\mu}$ is studied for the first time using $7.93~\mathrm{fb}^{-1}$ of data collected at $\sqrt{s}=3.773$ GeV with the BESIII detector. Benefiting from the much larger dataset compared to the previous experiment~\cite{plb_67-77}, we perform the first amplitude analysis for the $D^0\rightarrow K^-\pi^0\mu^+\nu_{\mu}$ decay and observe an $S$ wave component with a fraction $f_{S {\rm wave}}=(5.76 \pm 0.35_{\rm stat}   \pm 0.29_{\rm syst})\%$, resulting in $\mathcal{B}[D^0\rightarrow (K^-\pi^0)_{S {\rm wave}}\mu^+\nu_\mu]=(4.223 \pm 0.268_{\rm stat} \pm 0.222_{\rm syst})\times 10^{-4}$. The dominant $P$ wave component is observed with a fraction of $f_{K^{*}(892)^-}=(94.24 \pm 0.35_{\rm stat} \pm 0.29_{\rm syst})\%$, leading to  $\mathcal{B}(D^0\rightarrow K^{*}(892)^-\mu^+\nu_\mu)=(2.073\pm0.039_{\rm stat} \pm 0.032_{\rm syst})\%$ after considering $\mathcal{B}[K^{*}(892)^-\rightarrow K^-\pi^0]=1/3$~\cite{pdg16}. This result is consistent with previous measurement~\cite{plb_67-77} and has improved in precision by a factor of 5 over the current world average~\cite{pdg16}.  Benefiting from the improved precision,  our result  disfavors the covariant quark model (CQM) and  the covariant confining quark model (CCQM) calculations  for the first time~\cite{fpb14_66401,prd96_016017}, while  it supports the so-called chiral unitary approach ($\chi$UA) and light-cone sum rules (LCSR) calculations~\cite{prr2_0431129,prd92_054038,Ijmp21_6125-6172}. 
Combining with the previously most precise measurement of ${\mathcal B}(D^0\to K^{*}(892)^-e^+\nu_e)$~\cite{prd99_011103}, the ratio of the BFs is $\frac{{\mathcal B}(D^0\to K^{*}(892)^-\mu^+\nu_\mu)}{{\mathcal B}(D^0\to K^{*}(892)^-e^+\nu_e)} = 1.020\pm0.030_{\rm stat}\pm0.028_{\rm syst}$. 
No evidence of LFU violation is found with current statistics. Furthermore, the most precise FF ratios of the $D^0\rightarrow K^{*}(892)^-\mu^+\nu_{\mu}$ decay are determined to be $r_V=1.37 \pm 0.09_{\rm stat} \pm 0.03_{\rm syst}$ and $r_2=0.76 \pm 0.06_{\rm stat} \pm 0.02_{\rm syst}$.
The results are consistent with the CQM,  the CCQM,  the light-front quark model (LFQM),  and the  LCSR calculations~\cite{fpb14_66401,Ijmp21_6125-6172,cqm_2000,lfqm_2012}, and  they disfavor the HM$\chi$T model (based on the combination of heavy meson and chiral symmetries) calculation~\cite{hmt_2005}. All predictions and measurement results are summarized in Table~\ref{tab:Theory}. Our high-precision measurements have effectively advanced the study of the dynamics of SL decays of charmed mesons in the nonperturbative region, providing stricter constraints for the development of QCD theory.

\begin{table}[tp!]
\small
\caption{Measured the BF and FF ratios of $D^0\to K^{*}(892)^-\mu^+\nu_\mu$, and compared them with theoretical calculations and previous measurements.
 }
\begin{center}
\scalebox{0.75}{
\begin{tabular}
{lccc} \hline\hline Theory      & $\mathcal{B}$(\%)        & $r_V$     &  $r_2$     \\
       	\hline
                LCSR~\cite{prr2_0431129, Ijmp21_6125-6172} &  $2.01^{+0.09}_{-0.08}$ & 1.39  &  0.60\\
	       $\chi$UA~\cite{prd92_054038}  & 1.98 & ...  &  ...\\
                CCQM~\cite{ fpb14_66401}  & 2.80 & {1.22 $\pm $ 0.24 \;\;} &  0.92$ \pm $ 0.18\\
		CQM~\cite{cqm_2000,prd96_016017}  &3.09 &   1.56 &  0.74\\
		LFQM~\cite{lfqm_2012}  & ... & 1.36  &   0.83\\	
	        HM$_\chi$T~\cite{hmt_2005}     & ...  & 1.60  &  0.50 \\
                \hline 
                 Experiments     & $\mathcal{B}$(\%)        & $ r_V$     &  $r_2$     \\
                 \hline   
	         BESIII~\cite{ prd99_011103}    & $...$  & 1.46 $ \pm $ 0.07 $\pm $ 0.02 &  0.67 $\pm $ 0.06 $\pm $ 0.01 \\	              
	         FOCUS~\cite{plb_67-77}     & 1.89 $\pm$ 0.24  & 1.71 $\pm$ 0.68 $\pm$ 0.34  &  0.91 $\pm $ 0.37 $\pm $ 0.10 \\
	         This work     & 2.073 $\pm $ 0.039 $\pm $ 0.032  & 1.37 $\pm $ 0.09 $\pm $ 0.03  &  0.76 $\pm $ 0.06 $\pm $ 0.02 \\
	          
\hline\hline
\end{tabular}
  }
\label{tab:Theory}
\end{center}
\end{table}

\input{acknowledgement_2023-12-13}

\clearpage

\end{document}

%% file: authorlist_2023-12-13.tex
\author{
\small
M.~Ablikim$^{1}$, M.~N.~Achasov$^{4,c}$, P.~Adlarson$^{75}$, O.~Afedulidis$^{3}$, X.~C.~Ai$^{80}$, R.~Aliberti$^{35}$, A.~Amoroso$^{74A,74C}$, Q.~An$^{71,58,a}$, Y.~Bai$^{57}$, O.~Bakina$^{36}$
% ...
\\
\vspace{0.2cm}
(BESIII Collaboration)\\
\vspace{0.2cm} {\it
$^{1}$ Institute of High Energy Physics, Beijing 100049, People's Republic of China\\
% ...
$^{3}$ Bochum  Ruhr-University, D-44780 Bochum, Germany\\
$^{4}$ Budker Institute of Nuclear Physics SB RAS (BINP), Novosibirsk 630090, Russia\\
% ...
$^{35}$ Johannes Gutenberg University of Mainz, Johann-Joachim-Becher-Weg 45, D-55099 Mainz, Germany\\
$^{36}$ Joint Institute for Nuclear Research, 141980 Dubna, Moscow region, Russia\\
% ...
$^{57}$ Southeast University, Nanjing 211100, People's Republic of China\\
$^{58}$ State Key Laboratory of Particle Detection and Electronics, Beijing 100049, Hefei 230026, People's Republic of China\\
% ...
$^{71}$ University of Science and Technology of China, Hefei 230026, People's Republic of China\\
% ...
$^{74}$ University of Turin and INFN, (A)University of Turin, I-10125, Turin, Italy; (B)University of Eastern Piedmont, I-15121, Alessandria, Italy; (C)INFN, I-10125, Turin, Italy\\
$^{75}$ Uppsala University, Box 516, SE-75120 Uppsala, Sweden\\
% ...
$^{80}$ Zhengzhou University, Zhengzhou 450001, People's Republic of China\\
\vspace{0.2cm}
$^{a}$ Deceased\\
% ...
$^{c}$ Also at the Novosibirsk State University, Novosibirsk, 630090, Russia\\
% ...
}
}

%% file: acknowledgement_2023-12-13.tex
\begin{acknowledgments}

The BESIII Collaboration thanks the staff of BEPCII and the IHEP computing center for their strong support. This work is supported in part by National Key R\&D Program of China under Contracts Nos. 2020YFA0406400, 2020YFA0406300; National Natural Science Foundation of China (NSFC) under Contracts Nos. 11635010, 11735014, 11835012, 11935015, 11935016, 11935018, 11961141012, 12025502, 12035009, 12035013, 12061131003, 12192260, 12192261, 12192262, 12192263, 12192264, 12192265, 12221005, 12225509, 12235017; Guangdong Basic and Applied Basic Research Foundation under Grant No. 2023A1515010121 and the Fundamental Research Funds for the Central Universities under Contract No. 020400/531118010467; the Chinese Academy of Sciences (CAS) Large-Scale Scientific Facility Program; the CAS Center for Excellence in Particle Physics (CCEPP); Joint Large-Scale Scientific Facility Funds of the NSFC and CAS under Contract No. U1832207; CAS Key Research Program of Frontier Sciences under Contracts Nos. QYZDJ-SSW-SLH003, QYZDJ-SSW-SLH040; 100 Talents Program of CAS; The Institute of Nuclear and Particle Physics (INPAC) and Shanghai Key Laboratory for Particle Physics and Cosmology; European Union's Horizon 2020 research and innovation programme under Marie Sklodowska-Curie grant agreement under Contract No. 894790; German Research Foundation DFG under Contracts Nos. 455635585, Collaborative Research Center CRC 1044, FOR5327, GRK 2149; Istituto Nazionale di Fisica Nucleare, Italy; Ministry of Development of Turkey under Contract No. DPT2006K-120470; National Research Foundation of Korea under Contract No. NRF-2022R1A2C1092335; National Science and Technology fund of Mongolia; National Science Research and Innovation Fund (NSRF) via the Program Management Unit for Human Resources \& Institutional Development, Research and Innovation of Thailand under Contract No. B16F640076; Polish National Science Centre under Contract No. 2019/35/O/ST2/02907; The Swedish Research Council; U. S. Department of Energy under Contract No. DE-FG02-05ER41374

\end{acknowledgments}

%% file: draft_D0tokpi0munu_resub_PRL_v3.bbl
\begin{thebibliography}{99}


       \bibitem{pr494_197} M. Antonelli  {\it et al.}, 
  \href{https://www.sciencedirect.com/science/article/pii/S037015731000133X?via%3Dihub}{Phys. Rep. {\bf 494}, 197 (2010).}
  
       \bibitem{prl10_531} N. Cabibbo,  
  \href{https://journals.aps.org/prl/abstract/10.1103/PhysRevLett.10.531}{Phys. Rev. Lett. {\bf 10}, 531 (1963).}
  
       \bibitem{ptp49_652}M. Kobayashi and T. Maskawa,
  \href{https://academic.oup.com/ptp/article/49/2/652/1858101}{Prog. Theor. Phys. {\bf 49}, 652 (1973).}
  
      \bibitem{prd52_2783}D.~Scora and N.~Isgur,
  \href{https://journals.aps.org/prd/abstract/10.1103/PhysRevD.52.2783}{Phys. Rev. D {\bf 52}, 2783 (1995).}
  
       \bibitem{prd44_3567}P.~Ball, V.~M.~Braun, and H.~G.~Dosch,
  \href{https://journals.aps.org/prd/abstract/10.1103/PhysRevD.44.3567}{Phys. Rev. D {\bf 44}, 3567 (1991).}
 
       \bibitem{fpb14_66401} M.~A.~Ivanov, J.~G.~K\"orner, J.~N.~Pandya, P.~Santorelli, N.~R.~Soni, and C.~T.~Tran,
  \href{https://link.springer.com/article/10.1007/s11467-019-0908-1}{Front. Phys. (Beijing) {\bf 14}, 64401 (2019).}
  
       \bibitem{Ijmp21_6125-6172} Y.~L.~Wu, M.~Zhong, and Y.~B.~Zuo,
  \href{https://doi.org/10.1142/S0217751X06033209}{Int. J. Mod. Phys. A {\bf 21}, 6125 (2006).}
 
        \bibitem{cqm_2000} D.~Melikhov and B.~Stech,
 \href{https://journals.aps.org/prd/abstract/10.1103/PhysRevD.62.014006}{Phys. Rev. D {\bf 62}, 014006 (2000).}
 
         \bibitem{lfqm_2012} R.~C.~Verma,
 \href{https://iopscience.iop.org/article/10.1088/0954-3899/39/2/025005}{J. Phys. G {\bf 39}, 025005 (2012).}
 
         \bibitem{hmt_2005} S.~Fajfer and J.~F.~Kamenik,
 \href{https://journals.aps.org/prd/abstract/10.1103/PhysRevD.72.034029}{Phys. Rev. D {\bf 72}, 034029 (2005).}

          \bibitem{plb_67-77} J.M. Link {\it et al.} (FOCUS Collaboration),
\href{https://www.sciencedirect.com/science/article/pii/S0370269304017241?via%3Dihub}{Phys. Lett. B {\bf 607}, 67 (2005).}
  
            \bibitem{ARNPS73_285-314} B.~C.~Ke, J.~Koponen, H.~B.~Li, and Y.~Zheng,
  \href{https://www.annualreviews.org/doi/10.1146/annurev-nucl-110222-044046}{Annu. Rev. Nucl. Part. Sci. {\bf 73}, 285 (2023).}
  
            \bibitem{NSR8_181} H.~B.~Li and X.~R.~Lyu,
  \href{https://academic.oup.com/nsr/article/8/11/nwab181/6381732}{Natl. Sci. Rev. {\bf 8}, nwab181 (2021).}
  
             \bibitem{prd91_094009} S.~Fajfer, I.~Nisandzic, and U.~Rojec,
 \href{https://journals.aps.org/prd/abstract/10.1103/PhysRevD.91.094009}{Phys. Rev. D {\bf 91}, 094009 (2015).}
    
            \bibitem{cpc_063107} X.~Leng, X.~L.~Mu, Z.~T.~Zou, and Y.~Li,
  \href{https://iopscience.iop.org/article/10.1088/1674-1137/abf489}{Chin. Phys. C {\bf 45}, 063107 (2021).}
  
            \bibitem{prr2_0431129} H.~B.~Fu, W.~Cheng, L.~Zeng, D.~D.~Hu, and T.~Zhong,
  \href{https://journals.aps.org/prresearch/abstract/10.1103/PhysRevResearch.2.043129}{Phys. Rev. Res. {\bf 2}, 043129 (2020).}
  
             \bibitem{prd92_054038} T.~Sekihara and E.~Oset,
  \href{https://journals.aps.org/prd/abstract/10.1103/PhysRevD.92.054038}{Phys. Rev. D {\bf 92}, 054038 (2015).}
  
             \bibitem{prd96_016017} N.~R.~Soni and J.~N.~Pandya,
  \href{https://journals.aps.org/prd/abstract/10.1103/PhysRevD.96.016017}{Phys. Rev. D {\bf 96}, 016017 (2017);}
   \href{https://journals.aps.org/prd/abstract/10.1103/PhysRevD.99.059901}{{\bf 99}, 059901(E) (2019)}.
     
             \bibitem{Lum_1} M.~Ablikim {\it et al.} (BESIII Collaboration),
   \href{https://iopscience.iop.org/article/10.1088/1674-1137/37/12/123001}{Chin. Phys. C {\bf 37}, 123001 (2013).}
   
             \bibitem{Lum_2} M.~Ablikim {\it et al.} (BESIII Collaboration),
  \href{https://www.sciencedirect.com/science/article/pii/S0370269315008990}{Phys. Lett. B {\bf 753}, 629-638 (2016);}
  \href{https://www.sciencedirect.com/science/article/pii/S0370269320307851?via%3Dihub}{{\bf 812}, 135982(E) (2021)}.
   
             \bibitem{Lum_3} M.~Ablikim {\it et al.} (BESIII Collaboration),
   \href{https://iopscience.iop.org/article/10.1088/1674-1137/ad70a0}{Chin. Phys. C {\bf 48}, 123001 (2024).}
     
              \bibitem{Ablikim:2009aa} M.~Ablikim {\it et al.} (BESIII Collaboration),
   \href{https://www.sciencedirect.com/science/article/pii/S0168900209023870?via%3Dihub}{Nucl. Instrum. Methods Phys. Res., Sect. A {\bf 614}, 345 (2010).}
   
              \bibitem{detector} K.~X.~Huang, Z.~J.~Li, Z.~Qian, J.~Zhu, H.~Y.~Li, Y.~M.~Zhang, S.~S.~Sun and Z.~Y.~You,
   \href{https://link.springer.com/article/10.1007/s41365-022-01133-8}{ Nucl. Sci. Tech. {\bf 33}, 142 (2022).}
 
              \bibitem{geant4} S. Agostinelli {\it et al.} (GEANT4 Collaboration),
   \href{https://www.sciencedirect.com/science/article/pii/S0168900203013688?via%3Dihub}{ Nucl. Instrum. Methods Phys. Res., Sect. A {\bf 506}, 250 (2003).}
   
              \bibitem{kkmc} S. Jadach, B. F. L. Ward and Z. Was,
  \href{https://www.sciencedirect.com/science/article/pii/S0010465500000485?via%3Dihub}{ Comput. Phys. Commun. {\bf 130}, 260 (2000);}
  \href{https://journals.aps.org/prd/abstract/10.1103/PhysRevD.63.113009}{Phys. Rev. D {\bf 63}, 113009 (2001).}
  
             \bibitem{evtgen} D. J. Lange,
  \href{https://www.sciencedirect.com/science/article/pii/S0168900201000894?via%3Dihub}{ Nucl. Instrum. Methods Phys. Res., Sect. A {\bf 462}, 152 (2001)}; 
  R. G. Ping, \href{https://iopscience.iop.org/article/10.1088/1674-1137/32/8/001}{Chin. Phys. C {\bf 32}, 599 (2008).}
  
              \bibitem{pdg16} R. L. Workman {\it et al.} (Particle Data Group),
 \href{https://academic.oup.com/ptep/article/2022/8/083C01/6651666}{Prog. Theor. Exp. Phys. {\bf 2022}, 083C01 (2022).}
 
              \bibitem{lundcharm} J. C. Chen, G. S. Huang, X. R. Qi, D. H. Zhang and Y. S. Zhu,
 \href{https://journals.aps.org/prd/abstract/10.1103/PhysRevD.62.034003}{ Phys. Rev. D {\bf 62}, 034003 (2000);}
R. L. Yang, R. G. Ping and H. Chen, \href{https://iopscience.iop.org/article/10.1088/0256-307X/31/6/061301}{ Chin.  Phys.  Lett.   {\bf 31}, 061301 (2014).}

              \bibitem{photos}E.~Richter-Was,
 \href{https://www.sciencedirect.com/science/article/abs/pii/037026939390062M?via%3Dihub}{ Phys.  Lett.  B {\bf 303}, 163 (1993).}
 
              \bibitem{prl121_171803} M. Ablikim {\it et al.} (BESIII Collaboration),
 \href{https://journals.aps.org/prl/abstract/10.1103/PhysRevLett.121.171803}{Phys. Rev. Lett. {\bf 121}, 171803 (2018).}
 
              \bibitem{prl123_231801} M. Ablikim {\it et al.} (BESIII Collaboration),
 \href{https://journals.aps.org/prl/abstract/10.1103/PhysRevLett.123.231801}{Phys. Rev. Lett. {\bf 123}, 231801 (2019).}
 
              \bibitem{plb241_278} H. Albrecht {\it et al.} (ARGUS Collaboration),
 \href{https://www.sciencedirect.com/science/article/abs/pii/037026939091293K?via%3Dihub}{Phys. Lett. B. {\bf 241}, 278 (1990).}
 
               \bibitem{prd111_222} M. Ablikim {\it et al.} (BESIII Collaboration),
 \href{https://journals.aps.org/prd/abstract/10.1103/PhysRevD.109.072003}{Phys. Rev. D {\bf 109}, 072003 (2024).}
 
              \bibitem{pr168_1926} N.~Cabibbo and A.~Maksymowicz,
 \href{https://journals.aps.org/pr/abstract/10.1103/PhysRev.137.B438}{ Phys. Rev. {\bf 137}, B438 (1965);}
 \href{https://journals.aps.org/pr/abstract/10.1103/PhysRev.168.1926}{{\bf 168}, 1926(E) (1968)}.
   
              \bibitem{prd46_5040} C. L. Y. Lee, M. Lu and M. B. Wise,
 \href{https://journals.aps.org/prd/abstract/10.1103/PhysRevD.46.5040}{ Phys. Rev. D {\bf 46}, 5040 (1992).}
 
              \bibitem{cpc_063101} H.~Zhang, B.~C.~Ke, Y.~Yu and E.~Wang,
 \href{https://iopscience.iop.org/article/10.1088/1674-1137/acc642}{ Chin. Phys. C {\bf 47}, 063101 (2023).}
 
          \bibitem{prd94_032001} M. Ablikim {\it et al.} (BESIII Collaboration),
\href{https://journals.aps.org/prd/abstract/10.1103/PhysRevD.94.032001}{ Phys. Rev. D {\bf 94}, 032001 (2016).}
 
        \bibitem{prd83_072001} P. del Amo Sanchez {\it et al.} (\textit{BABAR} Collaboration),
 \href{https://journals.aps.org/prd/abstract/10.1103/PhysRevD.83.072001}{Phys. Rev. D {\bf 83}, 072001 (2011).}
 
         \bibitem{prd99_011103} M. Ablikim {\it et al.} (BESIII Collaboration),
\href{https://journals.aps.org/prd/abstract/10.1103/PhysRevD.99.011103}{Phys. Rev. D {\bf 99}, 011103 (2019).}
 
              \bibitem{prd85_122002} M.~Artuso {\it et al.} (CLEO Collaboration),
 \href{https://journals.aps.org/prd/abstract/10.1103/PhysRevD.85.122002}{Phys. Rev. D {\bf 85}, 122002 (2012).}
 
              \bibitem{corss_section} M.~Ablikim {\it et al.} (BESIII Collaboration),
   \href{https://iopscience.iop.org/article/10.1088/1674-1137/42/8/083001}{Chin. Phys. C {\bf 42}, 083001 (2018).}

\end{thebibliography}
